\documentclass[english,aps,pre, reprint, twocolumn]{revtex4}
\usepackage[T1]{fontenc}
\usepackage[latin9]{inputenc}
\usepackage{amsmath}
\usepackage{amssymb}
\usepackage{graphicx}

\makeatletter
\@ifundefined{textcolor}{}
{%
 \definecolor{BLACK}{gray}{0}
 \definecolor{WHITE}{gray}{1}
 \definecolor{RED}{rgb}{1,0,0}
 \definecolor{GREEN}{rgb}{0,1,0}
 \definecolor{BLUE}{rgb}{0,0,1}
 \definecolor{CYAN}{cmyk}{1,0,0,0}
 \definecolor{MAGENTA}{cmyk}{0,1,0,0}
 \definecolor{YELLOW}{cmyk}{0,0,1,0}
}

\makeatother

\usepackage{babel}
\begin{document}

\preprint{\%This line only printed with preprint option}

\title{Single-shot wideband active microrheology of viscoelastic fluids using pulse-scanned optical tweezers}

\author{Shuvojit Paul}

\affiliation{Indian Institute of Science Education and Research, Kolkata }

\author{Avijit Kundu}

\affiliation{Indian Institute of Science Education and Research, Kolkata}

\author{Ayan Banerjee}
\email{ayan@iiserkol.ac.in}

\affiliation{Indian Institute of Science Education and Research, Kolkata }
\begin{abstract}
We present a fast active microrheology technique exploring the phase response of a microscopic probe particle trapped in a linear viscoelastic fluid using optical tweezers under an external perturbation. Thus, we experimentally determine the cumulative response of the probe to an entire repertoire of sinusoidal excitations simultaneously by applying a spatial square pulse as an excitation to the trapped probe. The square pulse naturally contains the fundamental sinusoidal frequency component and higher odd harmonics, so that we measure the phase response of the probe over a very wide frequency band in a single shot. We then determine the responses to individual harmonics using a lock-in algorithm, and compare the phase shifts to those obtained theoretically by solving the equation of motion of the probe particle confined in a harmonic potential in the fluid in the presence of a sinusoidal perturbation. We go on  to relate the phase response of the probe to the complex shear modulus $G^{*}(\omega)$, and proceed to verify our technique in a mixture of polyacrylamide and water, which we compare with known values in literature and obtain very good agreement. Our method ensures that any drifts in time are almost entirely ruled out from the data, with the added advantage of high speed and ease of use.
\end{abstract}
\maketitle

\section{introduction}

Microrheology has become a ubiquitous tool in the determination of mechanical properties of complex fluids, and reveals important information about the microscopic structures of the constituents of the fluid \cite{furst2017microrheology,mason1995optical,squires2010fluid}. This is especially important in the case of biological fluids for various biophysical measurements \cite{tassieri2010analysis,tassieri2008dynamics,tassieri2008self,
	watts2013investigating} to probe intra-cellular environments \cite{kollmannsberger2011linear,harrison2013modes}, etc.  The advent of optical tweezers, where an optically trapped microscopic probe particle embedded in a fluid reveals the fluid's mechanical properties in the micro-scale, has been especially useful for microrheology applications since the motion of the probe is directly governed by fluidic properties in its immediate vicinity. Typically, complex fluids are viscoelastic in nature and are often parameterized in terms of the frequency dependent complex shear modulus $G^{*}(\omega)$ whose real part $G'(\omega)$ quantifies storage (or the `perfect solid'-like behaviour of the liquid) and the imaginary part $G''(\omega)$ quantifies loss (the `perfect liquid'-like behaviour of the liquid) \cite{ferry1980viscoelastic, mason1995optical, doi1988theory, Tassieri}. Thus, in  micro-rheology, the complex shear modulus $G^{*}(\omega)$ is measured by tracking the probe trajectory, which, in the absence of an external perturbations, is nothing but Brownian motion that is a consequence of the viscoelastic nature of the fluid through the generalized Stokes-Einstein relations. This is referred to as passive micro-rheology \cite{squires2010fluid}. On the other hand, in active micro-rheology, the probe is perturbed by an external force and the motion of the particle is analyzed to calculate $G^{*}(\omega)$ \cite{neckernuss2015active}. Active micro-rheology  can actually determine rheological properties over a wide spatial and temporal range, and also investigate nonlinearities, etc. with better signal-to-noise ratio (S/N) in comparison to the passive technique \cite{robertson2018optical, paul2018two}. A disadvantage of the active method, however, is that it is typically slow compared to the passive one since it gathers information only at the particular frequency of excitation. To cover a wide frequency band, the external perturbation frequency should be changed continuously and $G^{*}(\omega)$ should be measured for each frequency. Thus, any drift in fluid properties over time may affect the data causing erroneous inferences.

In this paper, we come up with a solution of this problem by introducing a fast active microrheology method using optical tweezer. We create pulse-scanned optical tweezers by modulating the trapping beam by a square pulse and proceed to record the trajectory of a trapped particle. Now, as the square pulse contains sinusoidal (odd) harmonics of all frequencies along with the fundamental with each component being orthogonal to others, a single time series actually contains information corresponding to all harmonics. Even considering that the signal-to-noise characteristics of each harmonic may vary - the information elicited is over a rather wide frequency band. We first present an algorithm to extract the phase lags of the particle responses for all harmonics of the pulsed tweezers from a single data set, and proceed to experimentally validate this algorithm in a viscoelastic fluid. Following this, we double the trap stiffness and repeat the measurement, and demonstrate that the phase lag changes without altering the viscoelastic parameters. Therefore, we eventually obtain two independent equations for two stiffnesses which later can be solved to calculate the complex shear modulus. Clearly, this method precludes further repetitions of the experiment with different pulse frequencies, so that it is much faster in collecting data over a large frequency range compared to other active microrheology methods in the literature.  Our method also enjoys the obvious advantages of phase measurements over amplitude measurements which include immunity to electronic noise that is often problematic when small changes in amplitude need to be measured, and most importantly, does not require the calibration of the signal in displacement units. Note that the latter demands a precise sensitivity measurement of the detection system every time a measurement is performed.  We also check the consistency of the method by calculating the complex shear modulus $G^{*}(\omega)$ for a viscoelastic sample (polyacrylamide and water) and compare it to values existing in the literature. Additionally, we make sure that the frequency dependent viscosity approaches the dc viscosity of the sample (the value of the solvent or the purely viscous component) at large frequencies, which happens since the elastic component cannot respond at high frequencies \cite{paul2018free,grimm2011brownian,ferry1980viscoelastic,raikher2010theory}. We proceed to describe the theory behind our measurements.

\section{Theory}
Let us consider a spherical microscopic particle  confined in a harmonic potential in an over-damped environment of a linear viscoelastic fluid with the minimum of the potential being sinusoidally modulated. Assuming that the fluid is a stationary, isotropic continuum around the particle, the equation of motion describing the particle's position $x(t)\forall t$ in one dimension can be written as
\begin{equation}
m\ddot{x}(t)=-\int_{-\infty}^{t}\gamma (t-t')\dot{x}(t')dt' -k[x(t) - x_{0}(t)]+\xi (t)
\label{eq1}
\end{equation}
where $m$ is the mass of the particle, $k$ is the stiffness of the harmonic potential, and $x_{0}(t)$ is the instantaneous position of the potential minimum. The integral term in the right-hand side, including the generalized time-dependent memory kernel $\gamma(t)$, represents damping by the fluid. Thus, $\gamma(t)$ can be termed as the time-dependent friction coefficient of the viscoelastic fluid. $\xi (t)$, on the other hand,  represents the Gaussian-distributed correlated, random force on the particle which models the thermal motions of the surrounding fluid molecules. The correlation of noise is given by $\langle \xi (t)\xi(t')\rangle=2k_{B}T\gamma (t-t')$ where $k_{B}$ is the Bolzmann constant and $T$ is the temperature, which is consistent with the fluctuation-dissipation theorem. Now, as we are in the low Reynold's number regime, the momentum of the particle relaxes rapidly so that its effect is experimentally undetectable. Therefore, we neglect the inertial term from the Eq.~\eqref{eq1}. Further, we average over the noise since we are only interested in the response of the particle under external perturbation. Thus, the Eq.~\eqref{eq1} in the frequency domain can be effectively written as
\begin{equation}
-i\omega \gamma(\omega)x(\omega)+k x(\omega)=k x_{0}(\omega)
\label{eq2}
\end{equation} 
Now, decomposing $\gamma(\omega)$ into its real and imaginary parts as $\gamma(\omega)=\gamma'(\omega) + j\gamma''(\omega)$, and calculating the phase of the response of the particle with respect to the external forcing, we obtain,
\begin{equation}
\tan(\phi)=-\frac{\omega\gamma'(\omega)}{k+\omega\gamma''(\omega)}
\label{eq3}
\end{equation}
Note that, the negative sign indicates that the response lags in phase with respect to the perturbation $kx_{0}(\omega)$. Now, we tune the trap stiffnesses from $k_{1}$ to $k_{2}$, and using Eq.~\eqref{eq3} after ignoring the negative sign, we obtain two equations
\begin{align}
\label{eq4}
\frac{1}{\tan(\phi_{1})}=\frac{k_{1}}{\omega_{0}\gamma'} + \frac{\gamma''(\omega)}{\gamma'(\omega)}\\
\label{eq5}
\frac{1}{\tan(\phi_{2})}=\frac{k_{2}}{\omega_{0}\gamma'} + \frac{\gamma''(\omega)}{\gamma'(\omega)}
\end{align}
where $\phi_{1}$, $\phi_{2}$ are the phases corresponding to the stiffnesses $k_{1}$ and $k_{2}$, respectively, and $\omega_{0}$ is the frequency by which the potential minimum is modulated (driving frequency). Now Eqns.~\eqref{eq4} and \eqref{eq5} can be solved to calculate $\gamma'(\omega)$ and $\gamma''(\omega)$ which are following:
\begin{align}
\label{eq6}
\gamma'(\omega)&=\frac{k_{1}-k_{2}}{\omega_{0}\left(\frac{1}{\tan(\phi_{1})} - \frac{1}{\tan(\phi_{2})}\right)}\\
\label{eq7}
\gamma''(\omega)&=\frac{\gamma'(\omega)}{\tan(\phi_{1})} - \frac{k_{1}}{\omega_{0}}
\end{align} 
Further, for a spherical probe of radius $a_{0}$, the real and the imaginary parts of the complex shear modulus $G^{*}(\omega)$ of a linear viscoelastic fluid are related to the friction coefficient as 
\begin{align}
\label{eq8}
G'(\omega)=\frac{\omega\gamma''(\omega)}{6\pi a_{0}}\\
\label{eq9}
G''(\omega)=\frac{\omega\gamma'(\omega)}{6\pi a_{0}},
\end{align} 
We can calculate these quantities by incorporating the previously calculated values of $\gamma'(\omega)$ and $\gamma''(\omega)$ into the Eqns.~\eqref{eq8} and \eqref{eq9}. In addition, we  can also calculate the frequency dependent viscosity from the knowledge of $G^{*}(\omega)$ as \cite{grimm2011brownian,doi1988theory}:
\begin{align}
\label{eq11}
\left|\eta (\omega)\right|=\frac{\sqrt{G'(\omega)^{2} + G''(\omega)^{2}}}{\omega}
\end{align}

\section{Experimental design} 
The details of our experimental setup is provided in Refs.~\cite{Bera2017,2018arXiv180804796P,paul2017,paul2018two}, here we describe it briefly for completeness. We built an optical tweezers around an inverted microscope (Zeiss Axiovert.A1 Observer) with an oil immersion objective lens (Zeiss PlanApo 100x, 1.4 numerical aperture) and a semiconductor laser of wavelength 1064 nm which is tightly focused on the sample. A piezo-mirror is placed at the conjugate plane of the objective focal plane to modulate the focal spot which essentially mimics the trapping potential minimum. We employ a second stationary and co-propagating laser beam  of wavelength $780$ nm and very low power ($<5\%$ of the trapping laser power) to track the probe particle's position (detection laser), which we determine from the backscattered light that is incident on a balanced detection system. The balanced-detection-system together with a data acquisition card (NI USB-6356) record the position of the probe in a host computer. We prepare a sample chamber of dimension around $20~mm \times 10~mm \times 0.2~mm$ by  attaching a cover slip to a glass slide by a double-sided tape which contains our model viscoelastic fluid. The model fluid samples are prepared by mixing polyacrylamide (PAM, flexible polyelectrolytes, $M_{w}=(5-6)\times10^{6} $ gm/mol, Sigma-Aldrich) into water along with mono-dispersed spherical polystyrene probe particles of radii $1.5 ~\mu m$ in very low volume fraction ($\approx 0.1\%$) which we use as probes. For each viscoelastic sample, first, we trap a single probe around $30 ~\mu m$ away from the nearest wall to get rid of any surface effects and record their Brownian motion with sampling frequency $10~kHz$ over $10~s$, to measure the stiffness of the trap. Then, we modulate the trap center by a square signal of very small peak-to-peak amplitude ($110 ~nm$) - so that the particle always remains in the linear region of the trap, and small frequency ($3 ~Hz$) - so that the harmonics are closely spaced in frequency. We record the data for $60$ seconds with sampling frequency of $10~kHz$, simultaneously with the excitation. We determine the phase response of the probe at different harmonics of the square wave excitation using a lock-in amplifier. 
\begin{figure}[]
	\centering
	\includegraphics[scale=0.35]{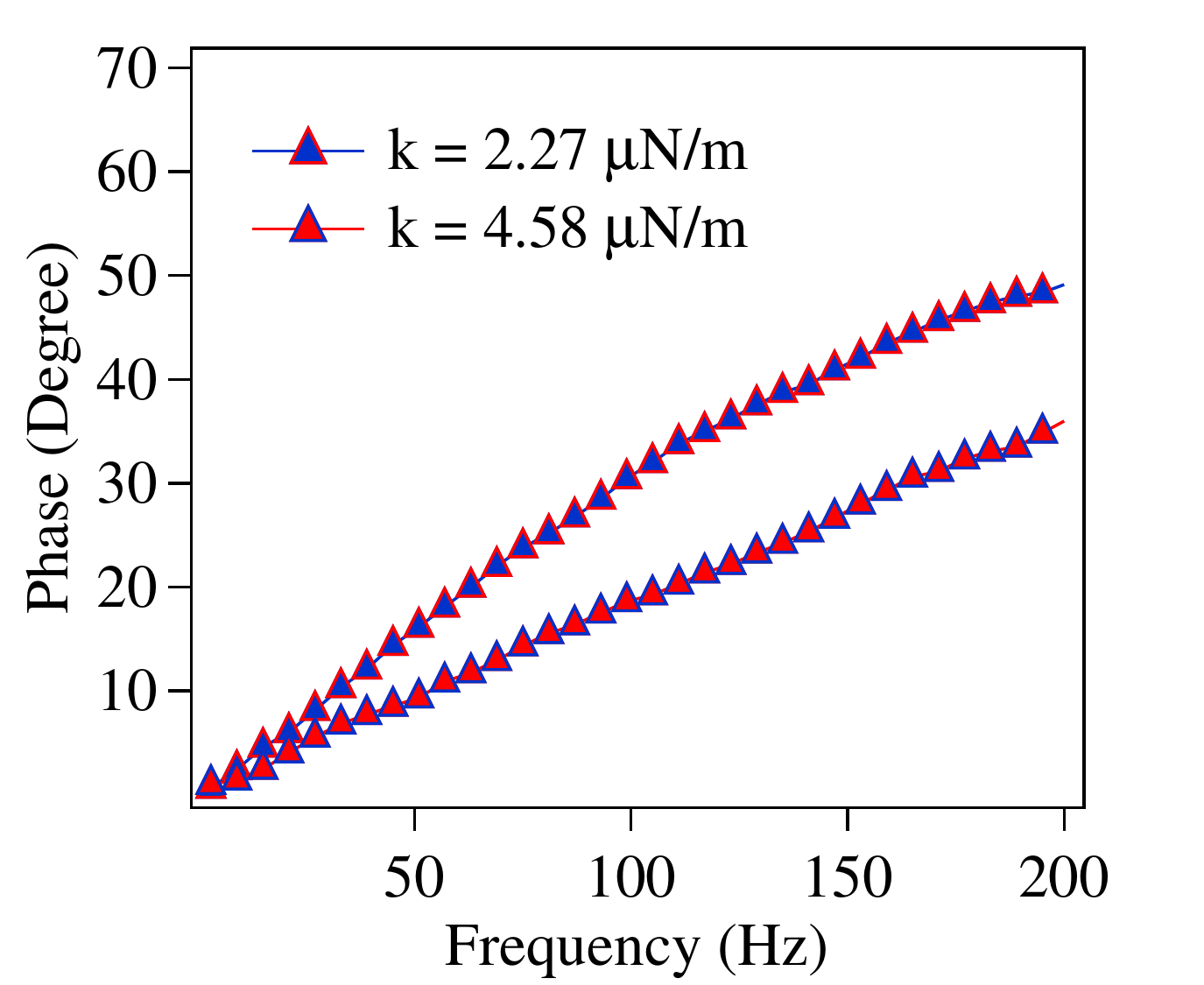}
	\caption{Phase lag of the probe response as a function of the sinusoidal components of the external square wave perturbation. The probe is embedded in PAM to water solution of $0.01~\%$ w/w .}
	\label{fig1}
\end{figure}

\section{Data analysis} 
We measure the stiffness of the trap using the equipartition theorem since it is independent of the rheological properties of the surrounding fluid. According to this theorem, the stiffness $k$ is related to the variance of the particle Brownian noise as $k=k_{B}T/\langle\left(x - \langle x \rangle\right)^{2}\rangle$ \cite{Tassieri,neuman2004kc}. For a second set of  measurements, we increase the trapping laser power and check that the stiffness changes to the double of its initial value. After the measurement of the stiffness, we modulate the trap center by a square signal and record the probe particle response. The square signal contains all the odd harmonics along with the fundamental with decreasing amplitude. Thus, the position of the trap center minimum $x_{0}(t)$ can be decomposed into a fourier series. Therefore, 
\begin{equation}
x_{0}(t)=\frac{4A}{\pi}\sum_{i=1}^{\infty}\frac{\sin(2\pi(2i-1)ft + \Phi_{i})}{2i-1}
\label{eq10}
\end{equation} 
where, $A$ is the amplitude, $f$ is the frequency of the fundamental mode and $\Phi_{i}$ is the modulation phase related to the i-th harmonic. Further, as all the harmonics are orthogonal to each other, the corresponding responses of the probe will be independent of each other. Now, let us assume that the response of the particle under the sinusoidal modulation is given by
\begin{equation}
x(t)=\sum_{i=1}^{\infty}A_{i}\sin(2\pi(2i-1)ft + \phi_{i}) + \zeta(t)
\end{equation}
where $A_{i}$ and $\phi_{i}$ are the amplitude and phase of the response corresponding to the i-th harmonic. $\zeta(t)$ is the noise corresponding to the Brownian fluctuations. We then multiply $x(t)$ by two computer generated references $\sin(2\pi(2i-1)ft)$ and $\cos(2\pi(2i-1)ft)$ and average over a large multiple of the reference time period, so that we have 
\begin{align}
X_{\sin}^{s}=\frac{A_{i}}{2}\cos(\phi_{i})\\
X_{\cos}^{s}=\frac{A_{i}}{2}\sin(\phi_{i})
\end{align}
respectively. It is because, all other terms will be zero. Now, going through the same procedure with the modulation, i.e., $x_{0}(t)$, we get
\begin{align}
	X_{\sin}^{m}=\frac{2A}{(2i-1)\pi}\cos(\Phi_{i})\\
	X_{\cos}^{m}=\frac{2A}{(2i-1)\pi}\sin(\Phi_{i})
\end{align}
respectively. Clearly thus,
\begin{align}
	\phi_{i}=\tan^{-1}\left(\frac{X_{\cos}^{s}}{X_{\sin}^{s}}\right)\\
	\Phi_{i}=\tan^{-1}\left(\frac{X_{\cos}^{m}}{X_{\sin}^{m}}\right)\\
\end{align} 
 and the relative phase between the signal and the modulation for the i-th harmonics is given by
 \begin{align}
 (\Delta\phi)_{i}=\phi_{i}-\Phi_{i}
 \end{align}  
 
\begin{figure}[]
	\centering
	\includegraphics[scale=0.42]{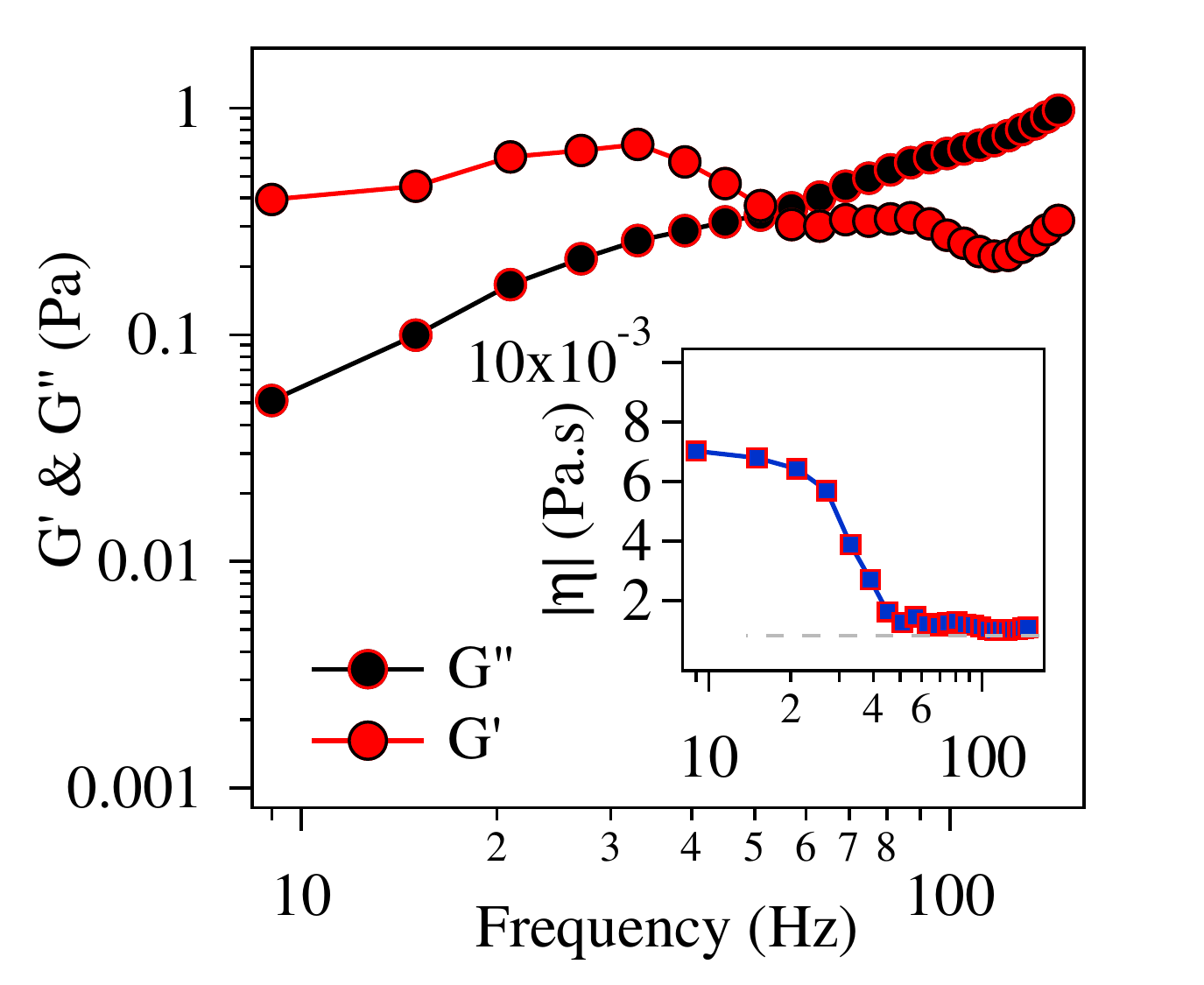}
	\caption{$G'$ and $G''$ as a function of (log) frequency for $0.01~\%$ w/w PAM to water solution. The inset shows the viscosity as a function of frequency, which at high frequencies, asymptotically approaches $0.00085~Pa.s$ - the viscosity of  water at $25^{\circ}$C.}
	\label{fig2}
\end{figure}

\begin{figure}[]
	\centering
	\includegraphics[scale=0.45]{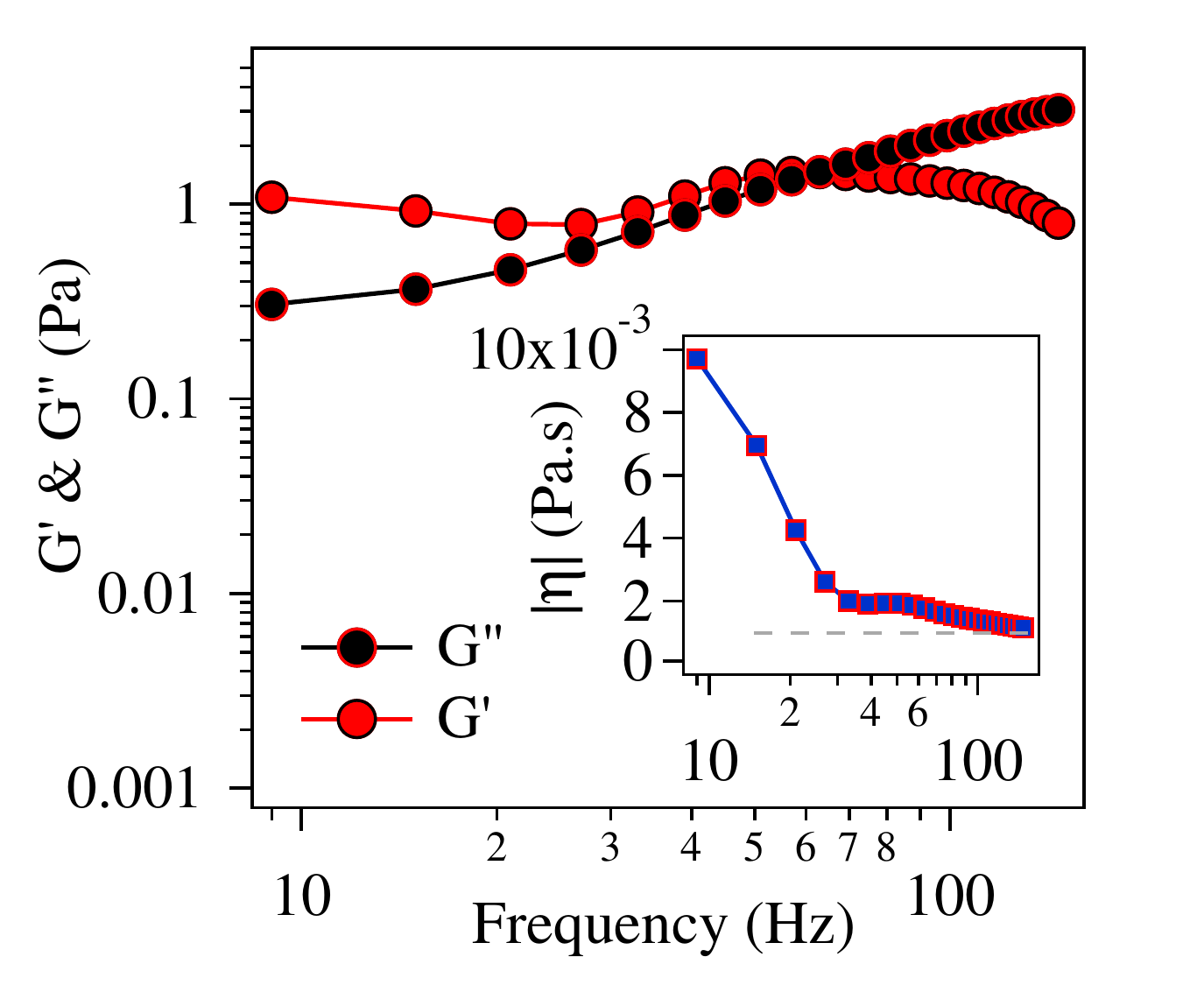}
	\caption{$G'$ and $G''$ as a function of (log)~frequency for $0.05~\%$ w/w PAM to water solution. The inset again shows that at higher frequencies, the viscosity asymptotically approaches the dynamic viscosity of  water at $25^{\circ}$C.}
	\label{fig3}
\end{figure}

 Therefore, by iterating over all the different harmonics of the reference signal, we can measure the phase lag corresponding to each harmonic in very minute time, and finally, we obtain the response of the probe over a wide frequency band from a single data set. It is thus no longer necessary to perform active microrheology at different excitation frequencies to obtain the probe response at each frequency, thereby increasing the total measurement time which may render it difficult to maintain identical ambient conditions. The signal-to-noise (SNR) of the phase lag measurements decrease with increasing frequency, and we consider an SNR of 3 as cut-off. In this, we can measure up to 50-66 harmonics of the square wave excitation frequency as shown in Figs.~\ref{fig1}-\ref{fig3}. Note that the SNR reduces for higher viscoelasticity samples which become more and more turbid, so that the backscattered signal itself becomes weaker, thus lowering the number of usable harmonics.

\section{Results and Discussions}
\begin{figure}[]
	\centering
	\includegraphics[scale=0.45]{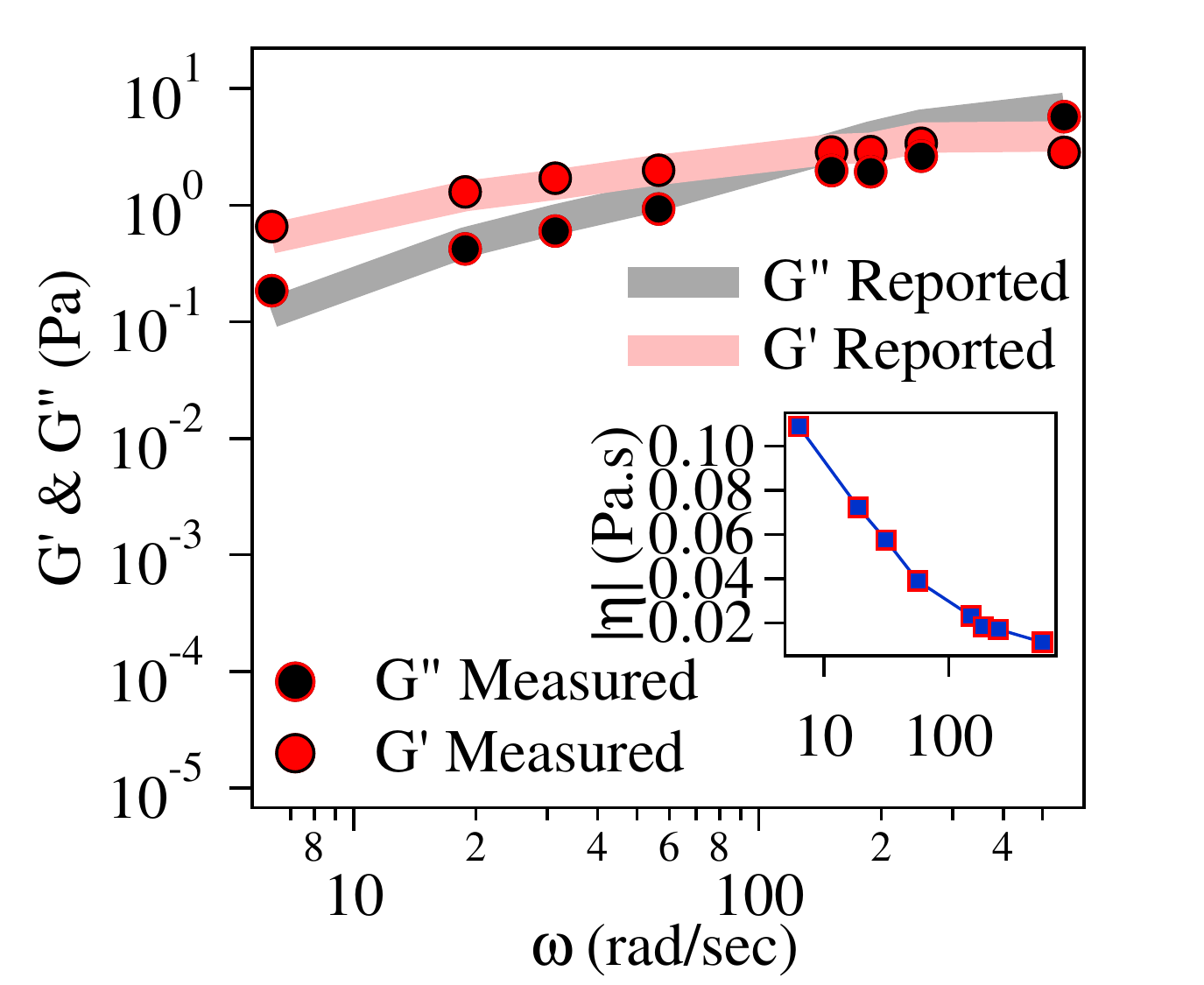}
	\caption{$G'$ and $G''$  as a function of (log) angular frequency for $1~\%$ w/w PAM to water solution compared to the results reported for the same concentration in  Fig.~5 of Tassieri et al. [Ref.~\cite{Tassieri}]). The inset shows the viscosity as a function of angular frequency which, as expected,  decreases with the increase of angular frequency.}
	\label{fig4}
\end{figure}

As we have described earlier, we calculate the trap stiffness from the equipartition theorem and kept the stiffness fixed throughout all our measurements for different fluid samples. Thus, the stiffness was  $2\pm 0.3~\mu N/m$ for the first laser power, and $4\pm 0.6~\mu N/m$ on doubling the power. Now for each sample, we measure the phase lag of the particle response for each laser power using the lock-in algorithm. We show the results for a linear viscoelastic fluid of $0.01~\%$ w/w PAM to water solution in Fig.~\ref{fig1}. Understandably, the phase lag is greater when the trap stiffness is low and it is less for higher stiffness of the trap. Now, using the measured phase responses and Eqns.~\eqref{eq6},\eqref{eq7},\eqref{eq8} and \eqref{eq9}, we proceed to calculate $G'$ and $G''$ for the samples. This we show in Fig.~\ref{fig2} for $0.01~\%$ w/w PAM to water solution. In the inset, we have plotted the frequency dependent viscosity against frequency which clearly approaches the viscosity of water at large frequencies since the elastic network in the fluid does not respond at these frequencies \cite{paul2018free,grimm2011brownian}. This serves as a good consistency check for our technique. Next, in Fig~\ref{fig3}, we show the real and imaginary parts of the complex shear modulus $G^{*}$ w.r.t frequency for $0.08~\%$ w/w PAM to water solution, with the frequency dependent viscosity in the inset. Here as well, the viscosity converges to $0.00085 ~Pa.s$ which is the viscosity of water at $25^{\circ}$ Celsius.  Now, a robust verification of our technique would be to compare our measurement of $G^{*}$ with that reported in the literature for the same sample. This we carry out for a $1~\%$ w/w PAM to water solution, and compare our measurement with results reported in Fig.~5 of Tassieri et al. [Ref.~\cite{Tassieri}], and obtained good agreement. We show this comparison in Fig.~\ref{fig4} where we plot the complex shear modulus $G^{*}$ against angular frequency for convenience of comparison with Ref.~\cite{Tassieri}. As the viscosity is too high for this sample, the amplitudes of the higher harmonics decrease extremely rapidly. Thus, we calculate $G^{*}$ for two fundamental frequencies $3~Hz$ and $8~Hz$, and eliminate points which are outliers from the general trend. Interestingly, in this case, we do not observe the viscosity reduce to that of water at high frequencies, though the inverse dependency does still remain. This may be due to the initial large value of viscosity, which causes a residual viscosity of the polymer network to remain at even high frequencies. Thus, we believe that our new method in active microrheology is robust and can be used to effectively determine $G^{*}$ even for viscoelastic solutions that have rather large values of the elastic component.

\section{Conclusions}

In conclusion, we report a wide-band active micro-rheology technique using pulse-scanned optical tweezers in which $G^{*}$ can be measured over a large frequency range for viscoelastic solutions of varying dilutions including those which have relatively large value of the elastic component. The method is fast since it obtains the frequency response of the medium over a wide range but simultaneously, or in a single shot, with a square wave excitation applied to an optically trapped probe particle embedded in the solution of interest. Thus, we simultaneously measure the phase response of the probe to a large number of sinusoidal (odd) harmonics present in the square wave using a lock-in algorithm implemented digitally. This mitigates the problem often encountered in active microrhelogy where probing the system at different frequencies is often compromised by a sudden change in the ambient conditions, which is not unlikely for the long measurement times naturally associated with the serial measurements at different excitation frequencies. To validate our method, we measure the storage and loss moduli, $G'(\omega)$ and $G''(\omega)$, respectively, of a linear viscoelastic sample and successfully check for agreement with values reported in literature.  We observe that we can measure the response continuously at harmonics 50-66 times higher than the fundamental square wave frequency depending on the viscoelasticity of the sample. In addition, we also show the expected inverse dependence of the viscosity on the excitation frequency, so that at large frequency values the viscosity approaches that of liquid water for most viscoelastic samples (PAM-water solutions of different w/w ratios), excepting the one with the highest PAM concentration. This establishes our method as robust and accurate, and indeed a new and fast approach in active microrheology. We intend to use this method to characterise diverse viscoelastic fluids, especially those associated with biological samples, and would finally like to extend it to the intracellular environment, where the complexity of the system and its constituents could lead to fascinating results. 
\begin{acknowledgments}
This work was supported by IISER Kolkata, an autonomous teaching and
research institute supported by the Ministry of Human Resource Development, Govt. of India.
\end{acknowledgments}

\bibliographystyle{apsrev4-1}
\bibliography{Fast_VE}

\end{document}